\documentstyle[12pt]{article}
\newcommand{\beq}{\begin{equation}}
\newcommand{\eeq}{\end{equation}}
\newcommand{\beqa}{\begin{eqnarray}}
\newcommand{\eeqa}{\end{eqnarray}}
\newcommand{\ba}{\begin{array}}
\newcommand{\ea}{\end{array}}

\makeatletter
\@addtoreset{equation}{section}

\makeatother

\begin{document}

\begin{titlepage}
\begin{flushright}
UT-Komaba/02-03

hep-th/0204202
\end{flushright}
\null
\vskip 1.5cm
\begin{center}

{\Large {\bf A Realization of $N=1$ ${\cal SW}(3/2,2)$ Algebras\\~\\
with Wolf Spaces}}

\lineskip .75em
\vskip 2.5cm
\normalsize

  {\large Michihiro Naka}

\vskip 1.5em

{\it Institute of Physics, University of Tokyo,\\
Komaba, Meguro-ku, Tokyo 153-8902, Japan}

\vspace{2ex}
              
{\texttt{hiro@hep1.c.u-tokyo.ac.jp}}

\vskip 6em

{\bf Abstract}
\end{center}

We find out that 
some unitary minimal models of 
the $N=1$ ${\cal SW}(3/2,2)$ superconformal algebra
can be realized as the level one coset models based on 
the Wolf spaces $SU(n)/\left(SU(n-2)\times SU(2)\right)$.
We obtain the expression of the fermionic current
with the conformal weight $5/2$ in the algebra.  
Then, these models 
are twisted to give the topological conformal field theories.
\end{titlepage}

\clearpage

\section{Introduction}
\hspace{5mm}
Study of representations of super ${\cal W}$ algebras 
is one of fundamental problems in two-dimensional
conformal field theories.
The super ${\cal W}$ algebras would
have their applications to provide an exact treatment of 
the superstring compactifications. 
Then we need a systematic understanding of the representations.
At present such study is quite difficult to overcome,
but Gepner and Noyvert \cite{GeNo} studied a construction 
of the unitary representations of
the simplest $N=1$ super ${\cal W}$ algebra, 
${\cal SW}\left(3/2,2\right)$ \cite{FiSc, KoMoNo}.
This super ${\cal W}$ algebra 
was expected to play a fundamental role in the string 
compactifications on $Spin(7)$ holonomy manifolds \cite{ShVa}.
Meanwhile, there was much progress on the study of the string 
compactifications  
on exceptional holonomy manifolds \cite{EgSu, BlBr, Noyvert, SuYa}.

It is well known that topological aspects of 
$N=2$ superconformal models are often relevant to
geometrical information on Calabi-Yau manifolds. 
Shatashvili and Vafa \cite{ShVa}
suggested a construction of
the topological twisting of the $c=12$ 
${\cal SW}\left(3/2,2\right)$ algebra,
and its implication to the conjectured mirror symmetry for 
$Spin(7)$ holonomy manifolds.
Their proposal is quite intriguing by itself
because it hints generic structure of the topological counterparts  
of the $N=1$ superconformal field theories
based on the super ${\cal W}$ algebras.
However, nobody has succeeded to provide an exact proof on the
topological twisting.
Motivated by this problem, 
we arrived at an observation on the 
$N=1$ ${\cal SW}(3/2,2)$ algebras with the central charges 
different from $c=12$.
The result arose from following two facts.

Let us begin with the result in \cite{GeNo} motivated by \cite{Romans}.
The $N=1$ ${\cal SW}\left(3/2,2\right)$ 
algebra consists of the two commuting Virasoro algebras, and
the fermionic spin $3/2, 5/2$ currents.
Then, we impose that one of the stress tensors 
should yield the $N=0$ minimal model.
This tightly
restricts possible central charges of the algebra.
The allowed values of the central charge are summarized in two discrete
series:
\begin{eqnarray}
c_p^{(1)}=6-\frac{18}{p+1},\qquad
c_p^{(2)}=6+\frac{18}{p},\nonumber
\end{eqnarray}
where $p$ is an integer such that $p\geq 3$.
I.e., $3/2 \leq c_p^{(1)}\leq 6$ and $6\leq c_p^{(2)}\leq 12$.
At these values, we have the unitary minimal model
which contain the representation of the 
$N=0$ minimal model of $c=1-6/p(p+1)$ 
as the subrepresentation of that of the $N=1$ algebra.
Then, the unitary model of $c_p^{(1)}$ has
the spin $3/2$ supercurrent which is primary  
with the conformal weight $h_{1,2}$ under 
the Virasoro current of the $N=0$ minimal model.
The conformal weights of the $N=0$ minimal model are given by
$h_{r,s}=(\left[rp-s(1+p)\right]^2-1)/4p(1+p),\,
(r=1\dots p,\; s=1\dots p-1)$.
Here, the point is that the fermionic spin $5/2$ current
of the $N=1$ algebra is descendant with respect to the 
spin $3/2$ supercurrent under the stress tensor 
of the $N=0$ minimal model.
Furthermore, it is possible to reconstruct  
the whole $N=1$ ${\cal SW}(3/2,2)$ algebra 
from these facts in the $N=0$ minimal model.
Closure of the operator product expansions of the $N=1$ algebra
determines the value of the central charge of the $N=1$ algebra. 
Subsequently, the unitary representations of the $N=1$ algebra
were determined. 

The other ingredient is the result in \cite{EgHoYa}.
We noticed that the central charge $c_p^{(1)}$
coincides with the central charge of the 
level one coset model based on the Wolf space 
$G/\left(H\times SU(2)\right)$ \cite{EgHoYa}:
\begin{eqnarray}
c=6-\frac{18}{g+1},\nonumber
\end{eqnarray}
where $g$ is the dual Coxeter number of the Lie group $G$
\footnote{
The other series of the central charge $c_p^{(2)}$
may be obtained by substituting 
the level of the current algebra into negative values formally.
We will not discuss this case in this note.}.
In fact,
we found out that the construction of the two Virasoro currents and
the $N=1$ spin $3/2$ superconformal current of the coset model
is the same as that of the $N=1$ ${\cal SW}(3/2,2)$ algebra in \cite{GeNo}.
This observation enabled us to write down 
the remaining fermionic spin $5/2$ current 
of the $N=1$ ${\cal SW}(3/2,2)$ algebra through the coset realization.
We will provide the explicit expression in the case with $G=SU(n)$,
but it should be straightforward to do with the other choices of $G$. 
Furthermore, we arrived at topological conformal field theories 
of the $N=1$ ${\cal SW}(3/2,2)$ algebra 
in the same way as \cite{EgHoYa, KaUcYa}.
The BRST-exact expression of the twisted 
stress energy tensor of the topological theories follows from 
the $c=0$ coset models of $\left(G\times G\right)/G$.
The BRST charge is given by the double contour integral of the
two spin $3/2$ supercurrents.
We expect that the structure of the topological twisting 
might shed some light on the suggestion in \cite{ShVa}.

Along these lines, we found out that the coset realization of the $N=1$ 
${\cal SW}(3/2,2)$ algebras via the Wolf spaces
$SU(n)/(SU(n-2)\times SU(2))$.
Then, we were led to the topological conformal field theories.
We will state these observations in section 2.
We will also include a relation of the $N=1$ coset models
to the $N=2$ coset models.
Section 3 will include conclusions and future problems 
arising from this note.

\section{Construction of the $N=1$ ${\cal SW}(3/2,2)$ Algebras}
\hspace{5mm}
We begin with the level one $SU(n)$ WZW theory 
using the Coulomb gas representation with the 
$n-1$ free scalar fields 
$\phi=(\phi_1,\phi_2,\dots,\phi_{n-1})$,
and choose its Coulomb gas parameters as: 
\begin{eqnarray}
\alpha_+=\sqrt{\frac{n+1}{n}},\qquad
\alpha_-=-\frac{1}{\alpha_+},\qquad
\alpha_0=\alpha_++\alpha_-
=\frac{1}{n\alpha_+}.\nonumber
\end{eqnarray}
We also introduce the highest root and the Weyl vector 
of $SU(n)$ :
\begin{eqnarray}
\theta=\alpha_1+\alpha_2+\dots\alpha_{n-1},
\qquad
\rho_{SU(n)}=\sum_{i=1}^{n-1}\frac{i(n-i)}{2}\alpha_i,\nonumber
\end{eqnarray}
where $\alpha_i \; (i=1, \dots, n-1)$ are the simple roots of the $SU(n)$.

Let us define the $SU(2)$ subgroup of $SU(n)$ by 
$E_{\theta}, E_{-\theta}$ and $[E_{\theta}, E_{-\theta}]$ 
where $E_{\theta}, E_{-\theta}$ are the lowering, raising operators.
This gives the subgroup $SU(n-2)$ whose root vectors 
span the orthogonal subspace to $\theta$ in the root space of $SU(n)$.
Then, we obtain the Wolf space $SU(n)/(SU(n-2)\times SU(2))$ which is
a symmetric space with a quaternionic structure \cite{Wolf}.
Let us decompose the Weyl vector according to the Wolf space:
\begin{eqnarray}
\rho_{SU(2)}=-\frac{\theta}{2},\qquad
\rho_{SU(n)}=\rho_{SU(n-2)}+(1-n)\rho_{SU(2)},\nonumber
\end{eqnarray}
which leads to the $N=1$ superconformal model of $c=6-18/(n+1)$ 
\cite{EgHoYa}.
Now, we will arrive at the $N=1$ ${\cal SW}(3/2,2)$ algebra. 
Let us recall that the generating currents of the algebra 
consist of the following four currents
\cite{FiSc, KoMoNo}:
the two bosonic spin $2$ stress tensors $T_{N=1}, T_{SU(2)}$, and
the fermionic spin $3/2, 5/2$ currents $G, U$.
The stress tensors are written down as 
\begin{eqnarray}
T_{N=1}(z)&=&-\frac{1}{2}(\partial\phi)^2(z)
+i\alpha_0\left(\rho_{SU(n-2)}+\rho_{SU(2)}\right)
\partial^2\phi(z),\nonumber\\
T_{SU(2)}(z)&=&-\frac{1}{2}\left(\sqrt{2}\rho_{SU(2)}
\partial\phi\right)^2(z)
+\frac{i}{\sqrt{2}}\alpha_0\left(\sqrt{2}\rho_{SU(2)}
\partial^2\phi\right)(z).\nonumber
\end{eqnarray}
Comparing with the construction in \cite{GeNo},
$T_{N=1}$ ({\rm resp.} $T_{SU(2)}$) is
the stress tensor of
the whole $N=1$ ${\cal SW}(3/2,2)$ algebra 
(resp. the $N=0$ minimal model).
The fermionic currents $G, U$ turn out to be
\begin{eqnarray}
G(z)&=&\psi_{SU(2)}\left(\psi_{SU(n-2),1}+\psi_{SU(n-2),2}\right)(z),
\nonumber\\
&=&e^{i\alpha_+\alpha_1\phi(z)}+e^{i\alpha_+\alpha_{n-1}\phi(z)},\nonumber\\
U(z)&=&
-\frac{2}{3}h_{1,2}
\psi_{SU(2)}\; \partial(\psi_{SU(n-2),1}+\psi_{SU(n-2),2})(z)\nonumber\\
&&\qquad+\left(1-\frac{2}{3}h_{1,2}\right)\left(\partial \psi_{SU(2)}
\right)\left(\psi_{SU(n-2),1}+\psi_{SU(n-2),2}\right)(z),\nonumber
\end{eqnarray}
where $h_{1,2}=(n+3)/4n$ is the conformal weight
in the Kac's table of the $N=0$ minimal model.
Here, we have introduced the following vertex operators
\begin{eqnarray}
&&\psi_{SU(2)}=e^{-i\alpha_+\rho_{SU(2)}\phi},\nonumber\\
&&\psi_{SU(n-2),1}=e^{i\alpha_+
\frac{\alpha_1-\alpha_2-\dots -\alpha_{n-1}}{2}\phi},\qquad
\psi_{SU(n-2),2}=e^{-i\alpha_+
\frac{-\alpha_1-\dots -\alpha_{n-2}+\alpha_{n-1}}{2}\phi}
\nonumber.
\end{eqnarray}
The complete 
expression of the currents of the $N=1$ ${\cal SW}(3/2,2)$ algebra
of $c_n^{(1)}=6-18/(n+1)$
through the Coulomb gas representation based on the Wolf space
$SU(n)/(SU(n-2)\times SU(2))$ is the main result in this note.

\vspace{2ex}

In fact, the coset models have their topological counterparts 
\cite{EgHoYa, KaUcYa}.  
The topological stress tensor is defined as
\begin{eqnarray}
T_{c=0}(z)
&=&T_{N=1}(z)-i\alpha_0 n \rho_{SU(2)}\partial^2\phi(z),\nonumber\\
&=&-\frac{1}{2}(\partial\phi)^2(z)+i\alpha_0\rho_{SU(n)}
\partial^2\phi(z).\nonumber
\end{eqnarray}
This topological stress tensor can be written in the BRST-exact form
\begin{eqnarray}
T_{c=0}(z)
=\{Q_{BRST},\; e^{-i\alpha_+\theta\phi}(z)\}\nonumber.
\end{eqnarray}
Here, the BRST charge $Q_{BRST}$
is defined as the double contour integral of the two
spin $3/2$ superconformal currents $G$
using a suitable path of the integration
\begin{eqnarray}
Q_{BRST}=\int\int dz \; dw\; G(z) G(w),\nonumber
\end{eqnarray}
and satisfies the nilpotency: $Q_{BRST}^2=0$.
It is possible to prove this nilpotency 
by checking poles in the integral.
The BRST-exactness of  
the topological stress tensor is also proven 
by inserting the screening charge 
$\int\, dz\, e^{i\alpha_+\alpha_i\phi(z)}$ in the integral.
The conformal weight
$h_{c=0}$ of the fermionic current $G$ (resp. $U$)
under the topological stress tensor $T_{c=0}$
is an integer, i.e., one (resp. two).
In \cite{ShVa}, the fermionic spin $5/2$ current $U(z)$ 
is suggested to be BRST-equivalent to the
anti-ghost field $e^{-i\alpha_+\theta\phi(z)}$:
$U(z)= e^{-i\alpha_+\theta\phi(z)}+\{Q_{BRST},*\}$.
We do not understand a role of the 
current $U(z)$ in the present topological model.

\vspace{2ex}

Finally, we discuss that 
the $N=1$ superconformal coset models 
have a relation to the standard $N=2$ 
superconformal coset models \cite{KaSu}
(The same argument was given in \cite{MiSu}.).
Let us explain it with the Wolf space $SU(4)/(SU(2)\times SU(2))$.
There is the cyclic ${\bf Z}_4$ symmetry 
in the root system of the affine Lie algebra
$\hat{su}(4)$:
$\left(\alpha_0,\alpha_1,\alpha_2,\alpha_3\right)
\to
\left(\alpha_3,\alpha_0,\alpha_1,\alpha_2\right)$
(The reader should not confuse the root $\alpha_0$ with
the Coulomb gas parameter.).
Then, the $N=1$ superconformal currents $T^{N=1}, G^{N=1}$ 
are mapped into the $N=2$ superconformal
currents $T^{N=2}, G^{\pm, N=2}$
of the coset model of the symmetric space 
$SU(4)/(SU(2)^2\times U(1))$:
\begin{eqnarray}
T^{N=1}_{SU(4)/(SU(2)\times SU(2))}
&=&
-\frac{1}{2}(\partial\phi)^2
+i\alpha_0 \left(-\frac{\alpha_0+\alpha_2}{2}\right)
\partial^2\phi,\nonumber\\
&=&
-\frac{1}{2}(\partial\phi)^2
+i\alpha_0 \left(\frac{\alpha_1+\alpha_3}{2}\right)
\partial^2\phi,\nonumber\\
&:=&T^{N=2}_{SU(4)/(SU(2)^2\times U(1))},
\nonumber\\
&&\nonumber\\
G^{N=1}_{SU(4)/(SU(2)\times SU(2))}
&=&e^{i\alpha_+\alpha_0\phi}+e^{i\alpha_+\alpha_2\phi},\nonumber\\
&=&e^{-i\alpha_+\theta\phi}+e^{i\alpha_+\alpha_2\phi},\nonumber\\
&:=&G^{-,N=2}_{SU(4)/(SU(2)^2\times U(1))}+
G^{+,N=2}_{SU(4)/(SU(2)^2\times U(1))}.
\nonumber
\end{eqnarray}
Here, the part of the vertex operators with
the imaginary root $\delta=\sum_{i=0}^{n-1}\alpha_i$ 
can be ignored in the operator product expansions of these currents 
due to the orthogonality:  $(\delta,\alpha_i)=0$.
Further, it is possible to write down the $U(1)$ current 
of the $N=2$ model 
\begin{eqnarray}
J^{N=2}=2i \alpha_0 \left(\rho_{SU(4)}-\rho_{SU(2)^2}\right)
\partial \phi.\nonumber
\end{eqnarray}
In this way, we arrive at the relation of the $N=1$ coset models 
to the $N=2$ coset models of the $SU(n)$ group
with the $SU(2)$ factor in the denominator. 
Here, we do not have a role of the fermionic current $U$
in the $N=2$ coset models.
Alternatively, it is straightforward to 
obtain another class of the $N=2$ coset models.
For example, to obtain the coset model of
the symmetric space $SU(4)/(SU(3)\times U(1))$,
we delete the simple roots $\alpha_1, \alpha_2$ of $SU(4)$ 
in the construction of \cite{EgHoYa}.

\section{Conclusion and Outlook}
\hspace{5mm}
We pointed out that
the $N=1$ superconformal
${\cal SW}(3/2,2)$ algebras 
with the specific central charge 
are realized by the Wolf spaces $SU(n)/(SU(n-2)\times SU(2))$.
Then, we discussed some consequences of this observation:
the existence of the topological conformal field theories 
and the relation to the $N=2$ superconformal coset models.
Through the realization, we confirmed the existence of 
the topological conformal
field theories based on the $N=1$ super ${\cal W}$ algebras.

\vspace{2ex}

Based on the result in this note, 
a lot of things should be considered:

\vspace{2ex}

(1) It is desirable to analyze in detail the spectrum in the
$N=1$ superconformal coset models.
We will dwell only on the states in the NS sector of the 
$c=3/2 \, (p=n=3)$ model.
The primary fields in the NS sector are labelled by 
the conformal weights $h, a$ under the stress tensors $T_{N=1}, T_{SU(2)}$.
Then, we arrive at the vertex operators 
$1, \, e^{i\alpha_0\alpha_1\phi},\,
e^{i\alpha_0(2\alpha_1+2\alpha_2)\phi}$. 
Here, we have imposed that 
the conformal weight $a$ should exist in the Kac's table of the 
Ising model of the stress tensor $T_{SU(2)}$.
We write these states as
$|0, 0\rangle,\, |1/16, 1/16\rangle,\,
|1/2, 0\rangle$
in the way that the state with the conformal weights $h, a$ is
denoted by $|a, h-a\rangle$
(In fact, these states coincide with those of the 
minimal model of $c_{p=3}^{(1)}=3/2$ in \cite{GeNo}.).
Meanwhile, when we move into the topological conformal field theory,
these states turn out to be the BRST invariant states
in the sense that  $Q_{BRST}|\Psi\rangle=0$
where $|\Psi\rangle$ is the state corresponding to the vertex operator.
Subsequently, 
we should proceed to the highest weight states in the Ramond sector
of the $N=1$ coset model.
Then, it would be very interesting to consider 
a geometrical interpretation of the spectrum through the Wolf space.
It also might be of use 
to visit the gauged WZW model on the Wolf space \cite{GaKe}.

\vspace{2ex}

(2) 
It would be very interesting to clarify a relation between the present 
result and the result in \cite{GeNo}.
The unitary spectrum in \cite{GeNo} has the continuous parts.
The $N=1$ coset model and the ${\bf Z}_2$ orbifold of the coset model
of the $N=2$ super $W_3$ algebra \cite{Romans}
should provide the special realization of the continuous spectra.
Then, this might suggest a new kind of mirror symmetry
because
it is likely that the Wolf spaces and the orbifolds of 
the hermitian symmetric spaces are geometrically different 
even if the corresponding $N=1$ superconformal field theories are the same.

\vspace{2ex}

(3) 
Extension of the result into the Lie groups other than $G=SU(n)$
is of interest (see \cite{KaUcYa}).
Along the quaternionic structure of the Wolf spaces,
it would be interesting to consider the extension with $G=Sp(n)$.
Then, it is natural to ask how to obtain the coset realization of the $N=1$
super ${\cal W}$ algebra through a Hamiltonian reduction
along the well-known story in the $N=2$ coset model \cite{Ito, KoMoNo}.
In a related direction,
the $N=1$ ${\cal SW}(3/2,3/2,2)$ algebra
was studied using the $\left(SU(2)\times SU(2)\right)/SU(2)$ coset model
motivated by an application to the string compactifications on $G_2$
holonomy manifolds \cite{Noyvert}.
The result might be understood through a
Hamiltonian reduction as in \cite{FeSe}.  

\vspace{2ex}

We hope to report further developments on these problems elsewhere.

\section*{Acknowledgements}
\hspace{5mm}
We thank T. Eguchi, T. Kawai, Y. Koga, K. Ito
and A. Yamaguchi for helpful conservations.
We are also grateful to D. Gepner and B. Noyvert for comments.
Especially, we wish to thank 
H. Kanno for collaborative discussions
and comments on the manuscript.
Present research is supported 
by Japanese Society of the Promotion of Science under 
Post-doctorial Research Program (No. 0206911).

\end{document}